\documentstyle[12pt]{article}
  \def\lsim{\lower.5ex\hbox{$\scriptstyle\buildrel < \over \sim $}}
  \def\gsim{\lower.5ex\hbox{$\scriptstyle\buildrel > \over \sim $}}
  \def\eps{\varepsilon}
\begin{document}
\begin{center}
{\large\bf Fractal Diffusion in Smooth Dynamical Systems\\
with Virtual Invariant Curves
}\\
\vspace {5mm}
{\it        B.V. Chirikov\footnote{Email: chirikov@inp.nsk.su} and
              V.V. Vecheslavov\footnote{Email: vecheslavov@inp.nsk.su}\\ [3mm]
                \it Budker Institute of Nuclear Physics\\
                630090 Novosibirsk, Russia}\\
\end{center}

\begin{abstract}
Preliminary results of extensive numerical experiments
with a family of simple models specified by the smooth canonical strongly chaotic
2D-map 
with global virtual invariant curves (VICs) are presented.
We focus on the statistics of the diffusion rate
$D$ of individual trajectories for various fixed values of the model 
perturbation parameters
$K$ and $d$. Our previous conjecture on the fractal statistics determined by
the critical structure of both the 
phase space and the motion is confirmed and studied in some detail.
Particularly, we have found additional characteristics of what 
we termed earlier the VIC diffusion suppression which is related to a new
very specific type of the critical structure.
A surprising example of ergodic motion with a "hidden" critical structure
strongly affecting the diffusion rate was also encountered.
At a weak perturbation ($K\ll 1$) we discovered a very peculiar diffusion regime
with the diffusion rate $D=K^2/3$ as in the opposite limit of strong 
($K\gg 1$) uncorrelated perturbation but, to the contrary, with strong
correlations and for a very short time only. As yet, we have no 
definite explanation
of such a controversial behavior.

\end{abstract}


{\bf 1. Introduction: virtual invariant curves}

In a 2D-map (2.1) we are going to study here the diffusion crucially depends
on the global invariant curves (GICs) which cut the 2D phase space (a cylinder,
see next Section) of the motion.
Even a
single such curve is sufficient to completely block the global diffusion (GD)
in the action variable along the cylinder. As is well known by now 
the existence of the GICs
 does depend not only on the perturbation strength
but also on its smoothness. It is convenient to characterized the latter
by the temporal Fourier spectrum of the perturbation. For analytical
perturbation the Fourier amplitudes decay exponentially fast. In this case
GD sets up if the perturbation 
$\epsilon\gsim\epsilon_{cr}$ exceeds some critical value.
Otherwise, the chaos remains localized within relatively narrow chaotic
layers of nonlinear resonances. As a result, the GD is either
completely blocked by GICs or the rate of the diffusion as well as the 
measure of its domain decay exponentially in parameter $1/\epsilon$ as
$\epsilon\to 0$ (the so-called Arnold diffusion, for a general review
see, e.g., \cite{1,2,3}).

By definition, the Hamiltonian of a smooth system has the power--law 
Fourier spectrum with
a certain exponent $\beta +1$ (see, e.g., \cite{4} and
references therein). In this case the GD is always 
blocked for some
sufficiently small perturbation strength $\epsilon < \epsilon_{cr}(\beta)$
provided the smoothness
parameter
$\beta > \beta_{cr}$ exceeds the critical value.
This is similar to the analytical Hamiltonian except that the critical perturbation
depends now on the Hamiltonian smoothness ($\epsilon_{cr}(\beta)\to 0$ as
$\beta\to\beta_{cr}$).

To the best of our knowledge, the most strong rigorous result reads:
$\beta_{cr} < 4$ for a 2D-map like in this paper (see \cite{5}).
However, a simple physical consideration \cite{4} leads to even smaller
value $\beta_{cr}=3$ which is still to be confirmed somehow, theoretically
or numerically. In any event, the smoothness of our model here $\beta =2$
is even less. 

Until recently, the behavior of dynamical systems in the opposite case
$\beta < \beta_{cr}$ of a poor smoothness remained rather vague. 
Even though the most numerical data
seemed to confirm the simplest behavior of some universal GD 
(see, e.g,, \cite{6}) a few counterexamples were observed too (see, e.g.,
\cite{7,8}). In the latter some trajectories remained within a certain
restricted part of the phase space for a fairly long computation time.
No clear explanation of these strange events has been given as yet.

Meanwhile, about 20 years ago (!) a number of mathematical studies revealed
various possibilities for the existence of GICs
in smooth systems with $\beta < \beta_{cr}$ (see, e.g., \cite{10,8,9}).
To us, the most comprehensive analysis of this problem
was given by Bullett \cite{9}
who rigorously proved a strange survival of infinitely many GICs
 amid a strong local chaos. 
Surprisingly, all these interesting results remain essentially unknown,
at least to physicists. Apparently, this is because the 
abovementioned mathematical papers were restricted (perforce !) to what
could be done rigorously that is to the invariant curves only without
any attempt to analyse the very interesting and important transport processes
such as diffusion.
This is still in reach to the physical analysis and numerical (or laboratory)
experiments only.
As a result, only after recent accidental rediscovery of GICs
 in chaos by Ovsyannikov \cite{11} (which are still
unpublished (!), see \cite{12,13} for the full text of Ovsyannikov's theorem)
the intensive physical studies of this interesting phenomenon have begun
\cite{12,13,14,15,16}.

Interestingly, both authors \cite{9,11} made use of exactly the 
same model for which a strange locked-in trajectory was observed 
still much earlier \cite{7}.
Apparently, this is because such a model (a particular case of our model
with parameter $d=1/2$, see next Section) is the simplest one possessing those
curious GICs (for discussion see \cite{15}).
Perhaps the main surprise was in that the GICs include the separatrices
of nonlinear resonances which were always considered before 
as ones destroyed first
by almost any perturbation. The principal difference is that now the invariant
curves, separatrices including, do exist for the special values of the system
parameters only (say $K=K_m$). 

Even though there are infinitely many such special values of the parameter,
and the infinitely many GICs for each of the value of which
just a single GIC does completely block the GD
the probability of the latter that is the measure of such $K$-values is
apparently zero. So, a principal question to be answered is:
what would be the behavior of that system for an {\it arbitrary} value of $K$?
In \cite{16} we conjectured that even though the set of $K_m$ is not generally
everywhere dense \cite{9} the density of this set is apparently rather high,
so that one may expect some change (presumably suppression) of the diffusion
for every $K$ value as compared to the "usual" (familiar) dynamical system.
In other words, we guessed that the structure of the phase space and of the
motion therein may be changed by the formation of GIC at a close
$K$ value even if at almost any $K$ there is actually no GIC.
This is why we call now such a neighbour-$K$ invariant curve the {\it virtual}
one (VIC) with respect to {\it any} $K$ \cite{16}.

Preliminary numerical experiments presented in \cite{16} did confirm 
our conjecture. This was done by the prompt computation of the average diffusion
rate $D(K)$ as a function of parameter $K$ in the domain with  GICs,
real or virtual ones. The experiments revealed a very strong suppression of the 
diffusion, up to many orders of magnitude, restricted only by the computation
time. But what turned out to be even more interesting was a very complicated
(apparently fractal) structure of the dependence $D(K)$.
This seems to be a result of a very complicated structure of the model phase
space itself. Preliminary, the latter looks like a so-called critical structure
(see, e.g., \cite{4}) but a rather specific one due to a forest of VICs.

In the present paper we begin studying this seemingly new type of such a structure.
Specifically, we start with investigation of the statistical properties
of the diffuson as one of the characteristic processes in the chaotic motion.

\vspace{5mm}

{\bf 2. Model: the same again}

\hspace*{\parindent}  
For reader's convenience we repeat below the discription of the model
which is the same as in \cite{15,16}. The model is specified by a map
in canonical variables action (momentum) $p$ - phase $x$:
$$
   \overline{p}\,=\,p\,+\,K f(x)\, , \quad \overline{x}\,=\,x\,+
   \overline{p}  \quad mod\ 1 \, .                             \eqno (2.1)
$$
where $K=\eps > 0$ is perturbation strength (not necessarily weak), and
"force" $f(x)$ has a form of antisymmetric ($f(-y)=-f(y),\ y=x-1/2$)
piecewise linear "saw" of period 1.
The phase space of the model is a cylinder: $0<x<1\,,\ -\infty < p < +\infty$.

As in \cite{15,16} we are going to actually consider a family of maps
with another parameter $d$ (see Fig.1 in \cite{15}) and the force
$$
  f(x)=\left\{
\begin{array}{ll}
      {2x/(1-d)}\, , & \mbox{if } |x| \leq {(1-d)/2}\, , \\
     -{2y/d}\, , & \mbox{if } |y| \leq {d/2}\, , \\
\end{array} \right.                                          \eqno (2.2)
$$
where $y=x-1/2$, and the second parameter $d$ ($0 \leq d \leq 1$)
is the distance between the two "teeth" of the saw
$|f(x)|=1$ at points $y=y_{\pm}=\pm d/2$. 
The most studied particular case of the family corresponds to $d=1/2$
when the two-teeth saw $f(x)$ is symmetric. In the limit $d=0$
the two teeth merge in one, and all the invariant curves are destroyed.
This was observed and explained in \cite{15} for $K>0$. In the opposite
case $K<0$ (which is equivalent to $K>0\,,\ d=1$) 
the dynamics of the model is completely different, and we will not consider
it in this paper (for a brief discussion see \cite{15}).
In our 2D-map (2.1) the GIC supports rotation
of phase $x$ around cylinder which bars any motion in $p$ over GICs.
Unlike this the local invariant curve (LIC), surrounding, for example, a domain
of regular motion (see, e.g., \cite{4} and Section 5 below), corresponds
to oscillation in phase $x$ which allows other trajectories to bypass
that obstacle.

The GICs, separatrices including, do exist 
in the whole interval $0 < d < 1$ but for special $K$ values only
\cite{9,15,16}. Particularly, the invariant curves are completely 
absent \cite{9}
for sufficiently large parameter
$$
   K\,>\,K_B(d)=\frac{2 d^2}{1+d}\, , \qquad   0 < d < 1 \, .  \eqno (2.3)
$$

If $K\gg K_B$ (see below) the physical quantity of the main interest for us, 
the diffusion
rate $D$, can be approximately calculated from the Fourier expansion of force
(2.2) (for detailes see \cite{16})
$$
 f(x)=\sum_{n \geq 1} \frac{f_n}{n^{\beta}} \sin(2\pi n x) \, ,
                                                              \eqno (2.4)
$$
where
$$
 f_n= -\frac{2}{\pi^2}\frac{\cos(n \pi) \sin(n \pi d)}{d(1-d)} \, , \qquad
       \beta=2 \, .                                           \eqno (2.5)
$$
Particularly, in the limit $d=0$
$$
 f_n= -\frac{2}{\pi} \cos(n \pi)  \, , \qquad \beta=1 \, .      \eqno (2.6)
$$
the smoothness parameter $\beta$ becomes less by one but the both values
are less than critical $\beta_{cr}=3$.

The calculation of the diffusion rate and other quantities are done using
the standard analysis of the nonlinear resonances and their interaction 
(overlap)
(see, e.g., \cite{1,2,3,16}). The calculation is especially simple if one can
neglect the variation of coefficients $|f_n|\approx const$ in (2.4).
This simplification is exact for $d=0$ (2.6), and remains reasonably accurate
\cite{16} for
$$
    K\,\gsim\, 3K_B\,=\,\frac{6 d^2}{1+d} \eqno (2.7)
$$
Then the diffusion rate is approximately given
by a very simple standard relation
$$
  D(K)\,=\,\frac{\overline{(\Delta p)_t^2}}{t}\, 
  \approx \frac{256}{\pi^5}\,K ^{5/2}\, \approx 0.57\,K^{5/2}  \eqno (2.8)
$$
where $t$ is the motion time in map's iterations, and
parameter $K\ll 1$ is assumed to be sufficienly small. The latter expression
in (2.8),
which we will use below, is the result of extensive numerical experiments
in \cite{6} confirmed also in \cite{16} for $K\lsim 0.1$ (see \cite{16} and 
next Section).

Notice that the dependence $D(K)\propto K^{5/2}$ is different from the usual,
or better to say, the simplest one $D(K)\propto K^{2}$. This is explained
by the dynamical correlation of motion which is determined by the frequency
of the phase oscillation on nonlinear resonances
$$
    \Omega_n\,=\,\sqrt{\frac{2\pi Kf_n}{n^{\beta -1}}}
    \approx\,2\sqrt{K}\,\approx\,\Lambda_n(K)\,\ll 1     \eqno (2.9)
$$
Here $\Lambda_n$ stands for the Lyapunov exponent charaterizing 
the local exponential instability
of the motion which is the main criterion for dynamical chaos. 
Notice that for $\beta =1$ both $\Omega_n$ and $\Lambda_n$
do not depend on the
Fourier harmonic number $n$.
The exact value of the Lyapunov exponent in the limit $d=0$ 
is given by
$$
   \Lambda\,=\,\ln{(1\,+\,K\,+\sqrt{2K\,+\,K^2})}\,\approx\,
   \sqrt{2K}\,\ll 1 \eqno (2.10)
$$
The latter expression is the approximation for small $K$ (cf. Eq.(2.9))
which is fairly well within the region of application of Eq.(2.8)
($K\lsim 0.1$) with accuracy $\sim 1\%$.
As the time in our model is discrete (the number of map's iterations)
both correlation characteristics, (2.9) and (2.10), must be small,
hence the above restriction on parameter $K$ too.

In the opposite limit $K\gg 1$ the correlation between successive $x$ values
are negligible, and one arrives at the "usual" relation for the diffusion rate:
$$
   D(K)\,=\,K^2 \int_0^1 f^2(x)\, dx =\frac{K^2}{3}
                                    \eqno (2.11)
$$
independent of the parameter $d$.
In intermediate region ($K\sim 1$) the correlation causes 
the decaying oscillation
(see \cite{6}) which is beyond the scope of the present paper.

\vspace{5mm}

{\bf 3. Diffusion without any invariant curves: averages and moments}

As was already mentioned above there are no invariant curves for $d=0$.
Moreover, the motion is ergodic that is of the simplest structure of
the phase space (cf. Section 4 below).
Therefore, this particular case is not of the main interest to us by itself.
Nevertheless, it is a good introduction to our central problem considered
in Section 6 below. A similar approach was taken in our previous paper
\cite{16}.

First, we consider the time dependence of the diffusion rate $D(K;t)$.
The semicolon here instead of usual comma is intended to emphasize that this
time dependence is not a real physical contribution to the diffusion but rather
a combination of two different processes: the proper diffusion via accumulation
of random perturbation effects and a stationary regular oscillation of the 
diffusing variable ($p$ in our case) which is a sort of background for the
diffusion. Such a phenomenon can be roughly represented by a simple relation:
$$
   D(K;t)\,\sim\,D_{\infty}(K)\,+\,{B(K)\over t}   \eqno (3.1)
$$
where $B(K)$ is some function of the perturbation
(see, e.g., \cite{16} and Eq.(3.5) below). 
In other words, in many cases, the present 
studies including, the nondiffusing stationary part can be separated from
the diffusing part which much simplifies the analysis of this complicated 
process. All this can be describe, of course, via the standard method of the
correlation of perturbation. However, this would lead to a much more intricate
theoretical relations and, besides, to much less information on the diffusion 
dynamics (see, e.g., \cite{6}).

An example of the diffusion kinetics is presented in Fig.1. The computation 
was done as follows. A number of trajectories $M\gg 1$ with random initial 
conditions homogeneously distributed within the unit area of the phase cylinder
($0\leq x_0 < 1\,,\ 0\leq p_0 < 1$) were run for a sufficiently long time
with successive outputs at certain intermediate moments of time $t$ as shown
in Fig.1. Remember that $t$ is measured in the number of map's iterations.
Each output includes the diffusion rate $<D>$ averaged over all
$M$ trajectories, and the dimensionless variance of that
$$
   V_M\,=\,\frac{<D^2>\,-\,<D>^2}{2<D>^2}  \eqno (3.2)
$$
For Gaussian distribution of action $p$ this variance would be unity. 
This is indeed the case for a sufficiently long motion time when the measured 
diffusion rate reaches its asymptotic value $D_{\infty}$ (3.1). 
A quite different dependence $V_M(t)$ for previous smaller time is of no 
surprise (nor is it very interesting) as over there $D(t)$ depends 
on a completely different physical
process one needs to pass over.

A real surprise was the very beginning of the diffusion, the plateau in Fig.1.
This looks as a real diffusion unlike the following part of a stationary
oscillation. Moreover, the diffusion rate on the plateau $D_0=K^2/3$
is the maximal one (2.11) as for large $K\gg 1$. 
Another interesting observation is the duration of this strange diffusion
$$
   t_0\,\approx\,{1\over\Lambda}\,\approx\,{1\over\sqrt{2K}} \eqno (3.3)
$$
which is close to the inversed Lyapunov exponent, the rise time of the local
exponential instability of the underlaying chaotic motion. 
The last but not least curious property is the fast increase of variance (3.2):
$$
   V_M(t)\,\approx\,{t\over 3}\,, \qquad 
   2\,\leq\,t\,\lsim\,t_0 \eqno (3.4)
$$
as shown in Fig.1. 
This is qualitatively different from the behavior of the same diffusion rate
for large $K\gg 1$ when the variance $V_M\approx 1$ is usual. 
The dynamical mechanism of this strange transitional diffusion 
is not completely
clear and requires further studies. Apparently, it is related somehow to the
main correlation (2.9) on the dynamical scale (3.3). 
Even though the initial "diffusion" is relatively fast it goes on a short time
only so that the relative change of the initial distribution 
of trajectories
$|\Delta p|/|\Delta p|_0\sim \sqrt{D_0/\Lambda}\sim K^{3/4}\ll 1$ 
is negligible  
for small $K\ll 1$ unless the initial distribution
$|\Delta p|_0\lsim K^{3/4}$ is very narrow.
However, in the latter case the dependence $D(t)$ is very sensitive 
to the form of the initial distribution in $p$ as a few our preliminary
numerical experiments reveal. The variance of $D(t)$ is especially strong
for small $t\sim t_0$ in the region of that mysterious plateau but eventually
decays as $t\to\infty$ when the diffusion approaches its limit value 
$D_{\infty}$.
Apparently, this is related to a complicated
fine structure of the phase space and/or of the motion correlations.
This interesting question certainly
deserves further studies but in the present paper we consider the simplest,
homogeneous, distribution of the trajectory initial conditions on the phase 
cylinder.

In this particular case, 
a very simple and surprisingly accurate empirical relation for the
diffusion time dependence has been found starting from a qualitative picture
(3.1). It has the form:
$$
   D(t)\,\approx\,\frac{D_0\,+\,\tau\,D_{\infty}}
   {\left(1\,+\,\tau^{\gamma}\right)^{1/\gamma}}\,, \qquad \tau\,=\,c\Lambda\,t
   \eqno (3.5)
$$
Here $\tau$ is the dimensionless time with an empirical fitting parameter
$c\approx 1$ which is very close to one. The second empirical parameter
$\gamma\approx 4$ is less definite but it affects the turn of the dependence
$D(t)$ at $\tau\approx 1$ only.
This relaxation of the diffusion rate has two time scales:
(i) the plateau $\tau_{pl}=1$ or $t_{pl}=1/c\Lambda\approx 1/\sqrt{2K}\gg 1$,
and (ii) the relaxation $\tau_R=D_0/D_{\infty}\sim 1/\sqrt{K}\gg 1$ or
$t_R\sim 1/K$ which is still much longer.
Interestingly, the usual diffusion spreading of a very narrow initial
$p$-distribution on the relaxation time scale
$$
   |\Delta p|_R^2\,=\,D_{\infty}t_R\,=\,D_{\infty}(D_0/D_{\infty})
   /c\Lambda\,=\,D_0/c\Lambda\,=\,|\Delta p|_{pl}^2
$$
is exactly equal to the spreading on the plateau. Hence, the full relaxation
spreading is twice as large which is also directly seen from the empirical
relation (3.5):
$$
   |\Delta p|_R^2\,=\,D(\tau_R )\cdot\frac{\tau_R}{c\Lambda}\,\approx\,
   \frac{D_0\,+\,D_0}
   {\left(1\,+\,\tau_R^{\gamma}\right)^{1/\gamma}}\cdot{\tau_R\over c\Lambda}\,
   \sim K^{3/2}\,\ll 1
$$
and which is still much less than the unit $p$-period.

In Fig.1 the empirical relation (3.5)
is presented and compared to the numerical data
in the dimensionless variables $\tau$ and $D^*=D/D_{\infty}$ where $D_{\infty}$
is the asymptotic ("true") diffusion rate (2.8).
In these variables the curves with various $K$ values are similar, and
converge in the limit $\tau\to\infty$. 

Another interesting scaling can be done as follows. Let us calculate the
diffusion rate $D_{\infty}(D(\tau ))=D_{th}$ from Eq.(3.5), and plot the ratio
of that to the true rate (2.8):
$$
   {D_{th}\over D_{\infty}}\,\approx\,\frac{D(\tau )\cdot 
   \left(1\,+\,\tau^{\gamma}\right)^{1/\gamma}\,-\,D_0}
   {\tau D_{\infty}}\,\approx\,1  \eqno (3.6)
$$
Then, within the accuracy of scaling (3.5) and of fluctuations, this ratio must be
always close to unity. This is indeed the case except the plateau ($t\lsim t_0$)
where the rate $D(\tau )$ is almost independent of $\tau$ (see Fig.1).

The next important statistical property are the fluctuations of the diffusion rate.
One characteristic of those is the dispersion of tracjectories which is characterized by the variance (3.2). If all the trajectories would be statistically
independent the dispersion of the mean diffusion rate were
$$
   \left(\frac{\Delta <D>}{<D>}\right)^2\,=\,\frac{2V_M}{M\,-\,1} \eqno (3.7)
$$
By construction, the trajectories are independent indeed with respect to their
initial conditions but not necessarily to the corresponding diffusion rate.
To check this we repeated the computation of diffusion $N$ times with new and
independent initial conditions, and then calculated the second (new) dimensionless
variance for the average diffusion rate:
$$
   V_N\,=\,\left(\frac{<<D>^2>_N}{<<D>>_N^2}\,-\,1\right)\cdot
   \frac{M\,-\,1}{2V_M}\,\approx\,1 \eqno (3.8)
$$
Again, if Eq.(3.7) holds true  the variance $V_N$ should be close to one.

The time dependence of both variances, $V_M(t)$ and $V_N(t)$,
is shown in Fig.1. Remarkably, their behavior is qualitatively
different. The first variance $V_M(t)$ depends on the distribution function of $p$ in the ensemble of trajectories while
the second variance $V_N(t)$ is affected by the statistical
dependence (or independence) among trajectories whatever their distribution
function. The results of
our numerical experiments presented in Fig.1 clearly demonstrate that the
distribution in $p$ quickly deviates from the Gaussian one during the diffusion
on plateau, and come back only in the limit $t\to\infty$ when the 
diffusion rate $D\to D_{\infty}$ approaches the asymptotic value without
any nondiffusing part. Unlike this, the trajectories remain statistically
independent during the whole process of the diffusion relaxation. 
We will come back to discussion of this interesting point in the conclusion
to this paper (Section 7). 

Now we turn to the most informative statistical characteristic, the distribution
function $f(D)$ of the diffusion rate.

\vspace{5mm}

{\bf 4. Diffusion without any invariant curves: the distribution function}

In the main part of our paper (Section 6) we shall be primarily interested
in the distribution tail $D\to 0$ of very low diffusion rate. 
The shape of this 
tail is known to be an important characteristic of the critical structure of
the motion (see, e.g., \cite{4}). First indications of such a structure 
in the presence of the virtual invariant curves have been observed in 
\cite{16}.
Here we continue these studies. 

Since the statistics of the far tail is always rather poor we make use, as in
\cite{16}, of a special version of the integral distribution
$$
   F(D)\,=\,\int_0^D f(D') dD'\,\approx\,\frac{j}{J} \eqno (4.1)
$$
the so-called "rank-ordering statistics of extreme events" (see, e.g., 
\cite{17}).
To this end the following simple ordering of the $D(j)$ values (events)
of the diffusion rate
is sufficient: $D(j+1) > D(j),\ j = 1,2,...,J$. 
Then the integral probability is approximately given by the ratio $j/J$
as shown in Eq.(4.1). 

In computation we typically ran $M$ trajectories by $N$ times 
(see Section 3),
so that the maximal number of events reached 
$J = M\times N = 10^4\times 10 = 10^5$. To obtain the lowest possible 
$D$ values
and minimize, at the same time, a rather big output
we ordered all the computed events but printed out much less of 
those
$J_0\ll J$ in such a way to get some all $J_1<J_0$ first (the smallest) $D_j$
while the rest were printed out in a logarithmic scale.
An example of such distribution is presented in Fig.2 for $K=0.001$ in 
variables 
$D^*=D/<D>$ and $F(D^*)=j/J$ where $<D>$ is some average diffusion rate 
(see below). The upper distribution corresponds to a rather long motion time
$t=10^4\gg 1/K$ when the mean diffusion rate 
is already very close to the limit $D_{\infty}$. For the lower distribution
$t=10$ is very short and corresponds to the plateau.

At least in the former case when the $p$ distribution is Gaussian 
(see Section 3)
the distribution 
$$
   f(D)\,=\,\frac{\alpha^\lambda}{\Gamma (\lambda)}\,D^{\lambda -1}\,
   {\rm e}^{-\alpha D} \eqno (4.2)
$$
is the so-called Pearson $\Gamma$-distribution with the two moments
$$
   <D>\,=\,\frac{\lambda}{\alpha}\,, \qquad (\Delta D)^2\,=\,
   <D^2>\,-\,<D>^2\,=\,
   \frac{\lambda}{\alpha^2} \eqno (4.3)
$$
which are the mean and variance, respectively. 
For Gaussian $p$-distribution the reduced variance (3.2) $V_M=1$ whence
$$
   \left(\frac{\Delta D}{<D>}\right)^2\,=\,\frac{1}{\lambda}\,=\,2 \eqno (4.4)
$$
and $\lambda =1/2$ independent of $\alpha$. 
If, moreover, we introduce the dimensionless diffusion rate 
$$
   D\,\to\,D^*\,=\,\frac{D}{D_{\infty}} \eqno (4.5)
$$
with average $<D^*>=1$, we obtain from Eq.(4.3)
$\alpha =\lambda = 1/2$, too.
Then, the new distribution becomes
$$
   f(D^*)\,=\,\frac{\,(D^*)^{-1/2}\,{\rm e}^{-D^*/2}}{\sqrt{2\pi}} 
$$
and
$$
   F(D^*)\,=\,\int_0^{D^*} f(D')\, dD'\,\to\,\sqrt{\frac{2}{\pi}\,D^*} 
   \eqno (4.6)
$$
where the latter expression gives the asymptotics $D^*\to 0$ we need.
This asymptotics is in a very good agreement with empirical data 
in Fig.2 even at $D^*\approx 0.1$ (!). For very small $D^*$ the accuracy of the
agreement is limited by the fluctuations due to a few remaining 
points.
The smallest $D^*=8.3\times 10^{-11}$ corresponds to the estimate
$D^*_{min}\sim 1/J^2=10^{-10}$.

Since the distribution $f(D^*)$ in (4.6)
is also Gaussian of $\sqrt{D^*}$ the integral $F(D^*)$ admits a very simple
approximation found in \cite{21}:
$$
   F(D^*)\,\approx\,\left\{
\begin{array}{ll}
   1\,-\,\frac{\exp{(-D^*/2)}}{\sqrt{D^*}\,+\,1}\,, & D^*\,>\,1/2 \\
   \sqrt{\frac{2D^*}{\pi}}\,, & D^*\,<\,1/2 \\
\end{array} \right. \eqno (4.6a)
$$
     The relative accuracy $|\Delta F/F|<0.05$ of that
      approximation
      is better than 5\%
      in the whole range of $F$.
      Actually, the accuracy is even much better except
      a narrow interval at $D^*\sim 1/2$.

Thus, the upper distribution in Fig.2, which describes the real diffusion at
sufficiently long motion time, is well in agreement with the available theory. 
This is no longer the case for the lower distribution on the plateau.
In itself, this is not a surprise as in the latter case, unlike the former one,
the measured diffusion rate is mainly determined by nondiffusive processes.
However, a very interesting feature of this nondiffusive distribution is 
in that
the exponent of the power-law tail remains exactly the same 
as if the $p$-distribution were again a 
Gaussian one. The simplest explanation, quite plausible to us,
is in that the far 
 tail still represents such a distribution
which is a part of the whole distribution according to our original picture
expressed by estimate (3.1). One immediate inference would be decrease
of the tail probability if we  still use the same variable $D^*=D/D_{\infty}$.
This is indeed the case according to the data in Fig.2 !

A more difficult problem is the quantitative estimate of the distribution shift
for the motion time $t\lsim 1/K$ when the ratio $<D^*>=<D(t)>/D_{\infty} > 1$.
This shift can be characterized either via the probability decrease by
$R_F$ times for a fixed $D^*$ or via the increase of $D^*$ itself by $R_D$
times for a fixed probability. Notice that on the tail $R_D=R_F^2$ due to
the square-root dependence (4.6). Characteristic $R_D$ seems to us more
preferable since it describes the shift not only of the tail but also
(qualitatively) of the whole distribution $F(D^*)$. 

After some playing with the data we have found the following 
empirical relation for the tail shift:
$$
   R_D(D_*)\,\approx\,D_*^a \eqno (4.7)
$$
where the new diffusion ratio
$$
   D_*(\tau )\,\approx\,\frac{D_0}{\tau D_{\infty}}\,+\,1 \eqno (4.8)
$$
and the fitted exponent $a =0.45$.

The philosophy behind this relation is following. We start with our original
picture of a combined diffusive/nondiffusive process (3.1) which is almost
our final choice (4.8).
However, at the beginning we seemed to improve the original relation
by inclusion of our surprising discovery, the plateau. Specifically, 
we tried to make use of Eq.(3.5) which is in a good agreement with the
empirical data for the dependence $D(t)$ (see Fig.1). Also, we have found
that it partly described the distribution $F(D)$ too, except on that mysterious
plateau ! Then, our final, so far, step was the change from (3.5) back to
a version of (3.1) in the form (4.8). 

How strange it may seem this did work with a reasonable accuracy as the
insert in Fig.2 demonstrates. The remaining question "why?" is still 
to be answered in farther studies. Actually, this is a general serious
problem of the dynamical mechanism underlaying the plateau formation and
statistics. 

Our empirical relation (4.7) can be represented in a different way. Namely,
instead describing the actual distribution tail shifted with respect
to the limiting asymptotics (4.6) we may introduce the scaled diffusion rate
$$
   D\,\to\,\frac{D}{R_D} 
$$
whence
$$
   D^*\,\to\,\frac{D^*}{R_D} \eqno (4.9)
$$
The result is shown in Fig.2 as a beam of 10 scaled distributions which are
scattered now around asymptotics (4.6).

\vspace{5mm}

{\bf 5. Diffusion amid virtual invariant curves: the Lyapunov exponents}

Above we considered a very particular and most simple limiting case of our
model (2.2) with parameter $d=0$. In this case the motion is ergodic
\cite{6} which greatly simplifies the problem under consideration.
Nevertheless, we obtained a number of new results which form
the firm foundations for further studies. 

The most important new feature of the motion for $d>0$ is the so--called
divided phase space of the system that is a mixture of both chaotic as well
as regular components of the motion. 
This is a typical structure of a few--freedom dynamical
system (see, e.g., \cite{4}).

First of all, we need to exclude the regular trajectories
from further analysis of the diffusion statistics.
The standard well known method to do this is simultaneous
computation for each trajectory of the so--called Lyapunov
exponent $\Lambda$ that is the rate of the local exponential
instability of the motion (see, e.g., \cite{1,2,3} and
references therein). 
In a 2D canonical (hamiltonian) map like our model (2.2) there are two
Lyapunov exponents whose sum is always zero: $\Lambda_1+\Lambda_2=0$.
For a chaotic trajectory one exponent, say, $\Lambda_1=\Lambda_+>0$
is positive while another one $\Lambda_2=\Lambda_-<0$ is negative.
As a result, according to the standard definition of the Lyapunov exponent
in the limit $t\to\infty$, any tangent vector $(dx,dp)$ of the linearized
motion approaches the eigenvector corresponding to $\Lambda_+>0$.

A simple well known procedure for computing $\Lambda_+$, we made use in the
present work too, is the following. For each of $M$ trajectories with random
initial conditions $x_0,p_0$ we chose the tangent vector $(dx,dp)$
of random direction and unit modulus: $d\rho^2=dx^2+dp^2=1$. Then both maps,
the main one and the second one linearized with respect to the main reference
trajectory $x(t,x_0,p_0),\  p(t,x_0,p_0)$ were run simultaneously
during some time $t$.
Finally, the {\it current} $\Lambda (t)$ was calculated from the standard
relation:
$$
   \Lambda (t)\,=\,\frac{<\ln{\rho (t)}>}{t}          \eqno (5.1)
$$
where the brackets denoted the averaging over $M$ trajectories.
Unlike the formal mathematical definition of $\Lambda$
in the limit $t\to\infty$, in numerical experiments 
the Lyapunov exponent $\Lambda (t)$ is always time dependent,
perforce. 

In Fig.3 a few typical examples of the $\Lambda$ distribution are depicted
for the number of events in (4.1) $J=M$ equal to that of trajectories while
the number of printed--out points $J_0=M'\leq M$ is less except the case $d=0$.
The simplest one is for ergodic motion ($d=0$). It has a form of almost vertical
step which derivative $dF/d\Lambda\sim 10^4$ is very narrow $\delta$-function.
Notice that regular chain of points along $F$ axis has no special physical
meaning but simply reflects a particular accepted type of the distribution
$F(\Lambda_j)=j/J$ (4.1) with integer $j$.
The mean $\Lambda$ depends only on $K$ (see Eq.(2.10))
but not on the initial conditions. This example in Fig.3 shows the 
empirical/theoretical ratio which is very close to unity as expected.

Two other examples correspond to the same $K=0.45$ and $M=10^4$ but different
motion time $t=10^4$ and $10^5$ iterations. Both distributions have the same
step at the largest $\Lambda$ which corresponds to diffusive components
(not necessarily a single one) of the
motion similar to the ergodic case. However, the most interesting part of the
former is the rest of the distribution  which represents a rich motion structure
contrary to a dull one in the ergodic motion. 

The largest (but again not the most interesting) part of this structure is related
to the distribution steep cut-off at small $\Lambda$. Comparison of the two
distributions for different motion time $t=10^4$ and $10^5$ shows that the $\Lambda$ values of the trajectories in this region
decrease with increasing time, approximately 
as $\Lambda\sim 1/t$. This would mean that all these trajectories are regular
(see Eq.(5.1)) because the tangent vector $\rho$ does not grow. The relative
number of such trajectories gives the total area of regular motion on system's
phase cylinder. In a particular example under consideration it amounts to 
$A_{reg}=3177/10000\approx 0.318\ (t=10^5)$. Generally, that value depends on
a particular cut--off border from above chosen (see arrow in Fig.3).
This delicate
experimental problem is considerably mitigated by a lucky feature of
$\Lambda$ distribution in our model, namely,
a relatively wide plateau of $F(\Lambda )$
immediately above the cut--off with only a few trajectories on it.
However, the statistical accuracy 
$$
  \frac{\Delta A_{reg}}{A_{reg}}\,\approx\,(M\cdot A_{reg})^{-1/2}
  \eqno (5.2)
$$
is typically much worse, and can be improved by increasing the number of trajectories (and the computation time) only. 

Another interesting feature of $\Lambda$ distribution in our model is a 
characteristic "fork" shape of the cut--off. This is a result of negative
$\Lambda$ for many regular trajectories. 
Such a peculiar representation is obtained by ordering $\Lambda (t)$ values
with their signs but plotting out the moduli $|\Lambda (t)|$ only.
Thus, the lower prong of the fork corresponds to negative $\Lambda (t)< 0$
while on the upper one $\Lambda (t)>0$ are positive. This is due to the 
complex--conjugate Lyapunov exponents which results in a strictly bounded
oscillation of the tangent vector $(dx,dp)$ in this case. 
However, the area (5.4) $A_{\pm}\approx 0.20 < A_{reg}\approx 0.318$ is noticeably
less than the total regular domains $A_{reg}$. The rest is filled with the
trajectories which are also regular but {\it linearly} unstable.
This means the linear in time growth of the tangent vector $\rho (t)\sim t$
so that $\Lambda (t)\to 0$ remains positive but is vanishing in the limit 
$t\to \infty$.
This is the so--called marginal local instability with both $\Lambda_{\pm}=0$
equal zero (for discussion see \cite{18}). 
A curious point is that this seemingly exceptional case becomes the typical one
in a nonlinear oscillator system due to the dependence of oscillation frequencies
on the trajectory initial conditions. 
In fact, the bounded $\rho$ oscillation producing negative $\Lambda (t)$ is
the exceptional case instead.
The origin of this peculiarity is in a piecewise {\it linear} force in our
model (2.2). As a result, the motion in the main (and the biggest for large $K$)
regular domain around fixed point $x=1/2,\ p=0$ is plainly a harmonic oscillation
with the frequency ($K < d$)
$$
  \Omega\,=\,\arccos{\left(1\,-\,\frac{K}{d}\right)}\approx 1.47 \eqno (5.3)
$$
which remains the same in the whole regular domain of area
$$
   A_{\pm}\,=\,\frac{2\pi K}{d}y_{\pm}^2\left(1\,-\,\frac{K}{2d}\right)
   \approx 0.20 \eqno (5.4)
$$
Here $y_{\pm}=x_{\pm}-0.5=\pm d/2$ is the position of two singularities of the force (see Eq.(2.2) and below) which restrict the size of the regular domain
surrounded by the limiting ellipse to which both lines of singularity
$y_{\pm}=d/2=0.25$ are tangent. This ellipse is determined by
the initial conditions
$$
  p_0\,=\,0, \quad x_0\,=\,0.5\,+\,y_{\pm}\left(1\,-\,\frac{K}{2d}\right)\,
  \approx\,0.5\,\pm\,0.185  \eqno (5.5)
$$
All the numerical values above correspond to $K=0.45$ and $d=1/2$.
Within the ellipse the motion of tangent vector obeys the same equation as the
main motion, the only difference being an arbitrary length $\rho$ of the tangent
vector (for details see \cite{3} and references therein). 

Coming back to Fig.3 notice the decrease of the measured area $A_{\pm}$ with
increasing motion time. This is explained by penetration of trajectories into a very complicated critical structure
at the chaos border surrounding each regular domain (for details see, e.g.,
\cite{4}). For the same reason, the direct measurement of the whole regular region
$A_{reg}\approx 0.40$ by a single chaotic trajectory as long as $10^9$ iterations
gives a noticeably larger value as compared to $A_{reg}\approx 0.318$ from $10^4$
trajectories for $10^5$ iterations each.


In all the curiosity of the $\Lambda (t)$ distribution in regular components of the motion our main interest in the present study is the intermediate region
between the regular cut--off at most small $\Lambda (t)\to 0$ and the chaotic
step at maximal $\Lambda$ independent of $t$. The distribution in this region
also does not depend on motion time and characterizes the proper critical
structure of the chaotic motion.
In an example in Fig.3 this structure is presented by a relatively small probability step $\Delta F\approx 0.06$ at $\Lambda\approx 0.03$.
In the next Section a few other examples will be given too.

\vspace{5mm}

{\bf 6. Diffusion amid virtual invariant curves: the Critical statistics}

In Fig.4 we present three characteristic examples of the effect of the critical structure on the diffusion statistics. The dashed curve shows the "unperturbed"
distribution $F(D^*)$ (4.1) of the normalized diffusion rate $D^*=D/D_{norm}$ where the
normalizing rate $D_{norm}$ to be chosen in each particular case (see below).
The term unperturbed means here the ergodic case $d=0$ without any invariant
curves and critical structure (Section 4, the problem of critical structure
in this case is not as simple as it may seem, see below and Section 7). 
In this case the normalizing rate
$D_{norm}=D_{\infty}$ is the true asymptotic diffusion rate (4.5). 

Now we are interested in the effect of the critical structure which typically
arises in a nonergodic motion with its barriers for the chaos, or chaos borders.
The latter are particular, and very important, case of an invariant curve which
is transformed into itself under system's dynamics. As was discussed already
above (Section 1) there are several different types of invariant curves (ICs).

One is well studied and rather familiar chaos border surrounding any domain
with regular motion. In this paper we call it the local invariant curve (LIC)
which does not block the global diffusion (GD) around such a domain.
An important property of LIC is the robustness which means that a small change
of the system, say, of a parameter $K$ or $d$ cannot distroy the LIC but may
only deform it slightly. This implies that LICs are always present in any divided
phase space. 

Here we are mainly interested in a different IC type, the global invariant curves (GICs). 
Each GIC cuts the whole phase--space
cylinder ($x\ mod\ 1$) of our model, and thus completely prevents 
GD in $p$. Such ICs are less known, especially the most surprising of them,
 the separatrix of a nonlinear resonance.
However, those GICss are not robust in the model under consideration (see \cite{9}),
being destroyed by almost any arbitrarily
small perturbation of the system, particularly by a change of even a single its
parameter. In other words, such GICs do exist for the special values, say, 
$K=K_m$ only. Even though there are typically infinitely many such special values
the probability to find a GIC in a randomly chosen system is zero.
This is why we are interested in a more generic situation when our model has no
GICs at all. Yet, the effect of those still persists in a certain domain around
each $K_m$ !
For this reason we call such GICs the virtual invariant curves (VICs) in analogy
with other virtual quantities in physics like, for example, virtual energy levels 
in qantum mechanics. 
Notice that unlike a GIC the VIC is robust and, hence, generic.

Both LICs and GICs produce the so--called critical structure of the motion
(see, e.g., \cite{4}) which is typically characterized by a power--law 
distribution of principal
quantities. The corresponding exponents $c_n$ are called the critical exponents.
Their values are shown in Fig.4 at the related distributions. 
Notice that the opposite is generally not true that is a particular power law
does not necessarily indicates any critical structure.
In our model this is just the case for the ergodic motion where the diffusion 
rate
distribution is also characterized by an asymptotic ($D\to 0$)
power law with exponent $c_0=0.5$ (see above and Section 7).
However, an important difference between ergodic and nonergodic dynamics is in that
all {\it critical} exponents in the latter case $c_n<c_0$ are less 
than (generally noncritical) ergodic exponent $c_0$.
This is the main physical result of our preliminary numerical experiments we can
present and discuss already right now (see Fig.4).

Let us start with the
distribution for $K=0.45$ (upper solid line) which is far in the region
without VICs (the border of this region
is at $K_B(d=1/2)=1/3$, see Eq.(2.3) above and \cite{16}).
However, the regular trajectories ($A_{reg}\approx 0.318$) together with LICs 
and the related critical
structure are present. As a result the distribution (with $D_{norm}=D_{\infty}$)
is well deviates from
the unperturbed one for ergodic motion with $d=0$.
This type of the critical structure in a relatively narrow layer around a LIC
is well studied by now (see, e.g., \cite{4}) including some deviation
of a typical distribution from a pure power law. The latter would mean the
exact scale invariance of the underlaying critical structure in both 
the system phase space
as well as its motion time.

The critical structure is described by the so--called renormalization group,
or renormgroup for brevity. On the other hand, the motion equations for any
dynamical system also form a certain (dynamical) group. Such a fundamental
similarity allows one to interpret the critical structure as a certain dynamics
which was called the {\it renormdynamics} \cite{19,4}. In this picture
the exact scale invariance with a pure power--law distribution corresponds
to the simplest, periodic, renormdynamics
even though the original dynamics may be the most complicated chaotic motion.
The resolution of this apparent paradox is in that the latter complexity of the
original dynamics is "transferred" to the dynamical infinitely dimensional space 
of the renormdynamics leaving behind the most simple renormdynamics itself
(sometimes !). 

The latter limit is most studied simply because it is the
simplest one. However, the generic case is just opposite that is a typical
renormchaos is also chaotic \cite{20,19}. Particularly, this implies 
a certain chaotic
oscillation of the characteristic distribution around some average power law.
This is just the case for the upper distribution in Fig.4 under discussion. 
It is characterized by the average critical exponent
$c_1=0.3$ with fluctuations $\sim (c'_1-c_1)=0.1$.
Such an interpretation of the critical structure in question is known to be
typical but not necessarily unique (see below).
The really unique property of this critical structure is the infinite power law,
with or without the fluctuations. The term "infinite" corresponds here to the
range of a renormdynamical variable $\ln{D}\to -\infty$ with unrestricted variation
even though the diffusion rate itself $D>0$ is strictly bounded from below.

This is no longer the case for a new type of critical structure which
we have encountered in our problem and which is produced by VICs (=robust GICs)
rather than by the robust LICs.
As was already explained above the principal difference between the two is in that
the VIC is not an invariant curve at all. In terms of renormdynamics it means
that VIC can mimic a GIC for relatively large $\ln{D}$ only.
This is clearly seen in Fig.4 in the upper part of the distribution with local
critical exponent $c_2=0.09$ and parameter $K=0.335$ (points).
Here we have taken $D_{norm}=10^{-6} < D_{\infty}\approx 2\times 10^{-5}$
much smaller than the true diffusion rate $D_{\infty}$. This shifts the whole
distribution to the right to avoid the overlapping with other
distributions. 
This value is slightly
above the border $K_B(1/2)=1/3$ (see Eq.(2.3)) where there is a lot
of VICs without any GIC. As a result the range of characteristic critical exponent
$c_2$ is very short: $\Delta \ln{D^{*}}\approx 5$ as compared with the total available
range $\approx 25$. The rest of the distribution remains fairly close to the
unperturbed one. This would mean the absence of the critical structute over there
or, at least, its sharp change at $\ln{D}\lsim 2$. In case of the former interpretation the renorm-motion stops in the latter region.\\ In turn, this would imply a "dissipative" rather than "Hamiltonian" renormdynamics.
Notice that the main part of the distribution is close but not identical to the
unperturbed one including a slight difference in the characteristic exponent.
Does it means a certain very slow renorm-motion remains a very interesting open
question yet to be studied.
Interestingly, the larger critical exponent $c'_2=0.45$ is also close to the local
critical exponent $c'_1=0.4$ in the region without VICs or GICs and was interpreted
above as a random fluctuation in the renormchaos. Is it really true still
remains unclear.

Finally, the third distribution in Fig.4 (lower solid line) is actually coincides
with the unperturbed distribution ($D_{norm}\approx D_{\infty}$)
even though it corresponds to the region 
with many VICs and strong suppression of the diffusion ($K=0.3294$, see Fig.3
in \cite{16}). A deviation for very small $D^*$ is due to a poor statistics
at this end. Notice that the coincidence of both distributions is
not only asympotic ($F\to 0$) but complete, including the opposite limit $F\to 1$. And this is in spite of a rather large regular region $A_{reg}\approx 0.581$.
The origin of this peculiarity for a particular $K$ value
remains unclear. One possibility is that, for some reasons,
the area of the critical structure at the chaos border around
this regular domain is unusually small. Examples of such a
peculiarity in different models are known (see \cite{22}). Actually, in the latter
work the critical structure was found to be unusually large but hidden.
In other words, the motion was ergodic but with strong correlations (cf. the
unusual diffusion rate (2.8) for $K\ll 1$ in ergodic system at $d=0$).
Coming back to this case in Fig.4 we can conclude that our "unperturbed"
power--law distribution with exponent $c_0=1/2$ (dashed line) may well represent
a peculiar critical structure related to the strong hidden temporal
correlations rather
than to a purely spatial geometry of the phase space.
If this is true, indeed, the correlation decay may well be
not a power--law one at all as is the case in model \cite{22}
where such a hidden decay is purely exponential (see Fig.6
over there). 

At last, let us mention another peculiarity of the critical
structure in question: all the critical exponents found
so far are less, if only a little, than "unperturbed" or
"hidden" one $c_0=1/2$. The physical meaning of this
universal inequality is in that the critical structure under consideration
does always increase the probability of very low diffusion
rate $D\to 0$. The general mechanism of this is known (see,
e.g., \cite{4}), and is explained by the "sticking" of a
trajectory within a complicated critical structure which
slows down the diffusion. 
Interestingly, that the sign of the sticking effect may be
opposite when the sticking accelerates the diffusion up to
the absolute maximum $D(t)\propto t$ of homogeneous diffusion
rate \cite{23,24}.

To summarize,
we see that our "simple" model in the present paper really reveals a great variety of critical structure
still to be farther studied and understood.

\vspace{5mm}

{\bf 7. Conclusion: A hidden critical structure ?}

In this paper we present some preliminary results of the numerical experiments
with a family of simple models specified by the smooth canonical 2D-map (2.1)
with global virtual invariant curves (VICs). As in our previous paper \cite{16}
we make use here of the same strongly chaotic model, and 
focus again on the statistics of the diffusion rate
$D$ which proves to be of a very complicated (apparently fractal) type
determined by the so--called critical structure of both the 
phase space and the motion (see, e.g., \cite{4}).
In previous paper \cite{16} we studied the statistics of the mean diffusion rate
$<D(K)>$ averaged over ensemble of trajectories with random initial conditions.
Our main result there was observation of very big and irregular fluctuations of the
dependence $<D(K)>$, and
a long and very
slowly decaying tail of $<D>$ distribution for $<D>\to 0$.
We termed the latter effect the {\it VIC diffusion suppression}.

In the present paper we continue studying this interesting phenomenon in more
details. To this end we turn from the statistics of averages $<D(K)>$ as a function
of model parameter $K$ to that of individual trajectories for a given $K$.
In principle, such an approach provides the most deep insight into 
a statistical problem.
As the main statistical characteristic we have chosen the integral distribution
$F(D)$ in the form (4.1) for a poor statistics at $D\to 0$.
Preliminary results of our extensive numerical experiments presented in Fig.4
confirm, indeed, our earlier conjecture on a critical structure underlaying
the fractal dependence $<D(K)>$ in \cite{16}, the true sign of such a structure
being various power--law distributions found.
Moreover, besides the familiar well known critical structure exemplified in Fig.4
by the case with parameter $K=0.45$ we did observe many cases of a rather different
structure as one with $K=0.335$. The principal difference of the latter is its
{\it finite} size in the structure variable $\Delta \ln{D}\lsim 5$.
The natural explanation of this difference is the following. First of all, the VIC
is not a true invariant curve like GIC. The latter completely blocks the GD while
the former can, at most, inhibit the diffusion only. 
The mechanism of inhibition is
known to be the sticking of a trajectory inside a very complicated critical
structure. In turn, the sticking is the stronger (longer) the smaller is
the spatial and/or the longer is the temporal scale of
the critical structure. But for the VIC structure both are strictly restricted.
On the other hand, such a restriction is the weaker the higher is the VIC density.
In the system under consideration the latter is rather large, and hence the
restriction leaves enough freedom for a strong suppression of GD for almost
any $K$. Moreover, since the critical exponent of the VIC structure is typically
very small (for example $c_2=0.09$ in Fig.4) the probability of large suppression
is high even for a short critical structure (cf. \cite{16} for a different
characteristic of this phenomenon). This slow decaying suppression probability
is well ascertained in our numerics but we have not, as yet, any theoretical
explanation of such a behavior.

Now we come, perhaps, to the most interesting result of our current studies.
How strange it may seem, this brings us to the apparently simplest case of our
model with $d=0$ when the motion is ergodic. Can it still reveal any structure
on the grounds that the distribution $F(D)$ is also a power law (Fig.4)?
That is the question! Certainly, it is not the case if, in addition, parameter
$K\gg 1$ is big, and the diffusion rate has the standard form (2.11): $D\propto K^2$.
But what if $K\ll 1$ is small? At least, the diffusion rate becomes qualitatively
different: $D\propto K^{5/2}$. What does it mean? Generally, nothing!
But in a particular case under consideration such dependence $D(K)$ can be, and actually was, derived \cite{16} from the resonance structure of the motion.
If the system were not ergodic (with divided phase space) this structure would be
clearly seen in the phase space. The question is what happens for ergodic motion
with the same dependence $D(K)$?
In \cite{16} we conjectured that some structure would still persist in the form
of the correlations which determine the diffusion rate that is in some "hidden"
form unseen directly in the picture of the motion in phase space.
An example of such hidden critical structure was found in \cite{22} (see Section 6
above). However, in that case a particular distribution function was exponential
rather than a power--law one (?).
Hence, the question is if this qualitative difference could depend on a particular characteristic of the critical structure?
Still another question arises from a very strange temporal behavior of the diffusion rate in the same "simple" case of ergodic motion for $d=0$. We mean
a "mysterious" plateau at the very beginning of diffusion 
under a weak perturbation ($K\ll 1$, see Fig.1). In this case the dependence
$D(K)=K^2/3$ is the same as in the opposite limit of strong 
($K\gg 1$) uncorrelated perturbation (?) but for a very short time only,
the shorter the stronger is the perturbation (?!).
Moreover, the correlations on the plateau are not only very large as in the
weak--perturbation limit $K\to 0$ but even increasing during the whole plateau
regime (see Fig.1, dashed lines for variances $V_M(\tau )$ (3.2)).
As yet, we have no definite explanation for
such a controversial behavior.
A discreet current conjecture is the following. The duration of plateau is
$\tau_{pl}\approx 1$, or $t_{pl}\approx 1/\Lambda\approx 1/\Omega$ (see Eq.(2.9)).
But the latter expression gives the period of phase oscillation on the critical
nonlinear resonance which determines the diffusion rate \cite{16}. Then, one
can image that this period does characterize not only the correlation decay,
as usual, but also the correlation uprise. Still, the invariable diffusion rate
over the whole plateau region is to be explained yet.

In any event, we are very curious to continue these most interesting studies!

\vspace{3mm}

{\bf Acknowledgements.} This work was partly supported by the Russia Foundation
for Fundamental Research, grant 01-02-16836.
We are grateful to Ms. L.F. Hailo for her permanent and very important assistance in computer experiments.


\newpage    \begin{center}  Figure captions  \end{center}

\begin{itemize}
\item[Fig.1] Diffusion relaxation $D^*(\tau )=D(\tau )/D_{\infty}\to 1$ 
             in model (2.2) with parameter $d=0$ (without invariant curves)
             is presented as a function of dimensionless time $\tau$ (3.5)
             for two values of $K=0.01$ (circles) and $3\times 10^{-5}$ 
             (crosses).
             Two smooth solid lines show the empirical relation (3.5) with
             two fitting parameters $c=1$ and $\gamma =4$.
             Dashed lines are variances $V_M(\tau )$ (3.2),
             and dotted lines show variances $V_N(\tau )$ (3.8) (see text).
             In the lower part the scaling (3.6) is presented reduced 
             by factor 10
             to avoid overlapping with other data.
             The full volume of empirical data was 
             $J = M\times N = 10^4\times 10 = 10^5$ (see text).
             
\item[Fig.2] Distribution function $F(D^*)$ (4.1) of the reduced diffusion 
             rate $D^*$ (4.5) in model (2.2) without invariant curves ($d=0$).
             Thick dashed straight line represents
             asymptotics (4.6) of the integrated $D$-distribution (4.2) 
             for the Gaussian
             $p$-statistics. Two lower wiggly lines correspond
             to large deviations from the latter: $D_*=42\ (K=10^{-3})$ and 
             $461\ (K=3\times 10^{-5})$ (see Insert).
             A group of 10 $D$-distributions in a large interval 
             ($10\leq D_*\leq 461$) are brought together using empirical 
             relation (4.9). Insert: shift factor $R_D$ (see text) vs. 
             deviation $D_*$ (4.8) for $K=10^{-3}$ (circles) and
             $3\times 10^{-5}$ (crosses); the straight line is empirical
             relation (4.7).
              
\item[Fig.3] Examples of distribution function $F(\Lambda )$ 
             of type (4.1) but for the 
             Lyapunov exponent in model (2.2) for $d=0,\ M=M'=80,\ 
             t=10^4$
             (the right-most step $F(\Lambda )$, ergodic motion), and for
             $d=1/2,\ M=10^4,\ M'=1000,\ t=10^4,\ 10^5$
             (nonergodic motion, see text); in all cases $K=0.45$.
             Horizontal line indicates 
             the total share
             $A_{reg}\approx 0.318$ of the motion regular components.
             Arrow at $\Lambda = 10^{-4}$ shows the lower border of chaotic
             trajectories chosen for further  analysis 
             (for $t=10^5$, see text).
             
\item[Fig.4] Three characteristic examples of the diffusion statistics 
             in the critical structure including virtual 
             invariant curves ($d=1/2$).
             Shown are integral distributions $F(D^*)$ (4.1) of the normalized
             diffusion rate $D^*= D/D_{norm}$ (see text).
             The numbers at curves are critical diffusion exponents $c_m$.
             The biggest one $c_0=1/2$ corresponds to the ergodic
             motion ($d=0$) without any critical structure (dashed curve).
             Two straight lines show the average ($c_1=0.3$) and local ($c'_1=0.4$)
             critical exponents for $K=0.45$ (solid line connecting
             500 values of $F(D^*)$). The distribution for $K=0.335$ with two
             local critical exponents ($c_2=0.09$ and $c'_2=0.45$) is presented
             by 300 points shifted to the right to avoid overlapping with two other
             distributions. 
             The third distribution (a solid line through 1000 points,
             $K=0.3294$) is surprisingly close to that in ergodic case 
             (dashed line). In all three examples $M=10^4,\ t=10^5$.

\end{itemize}

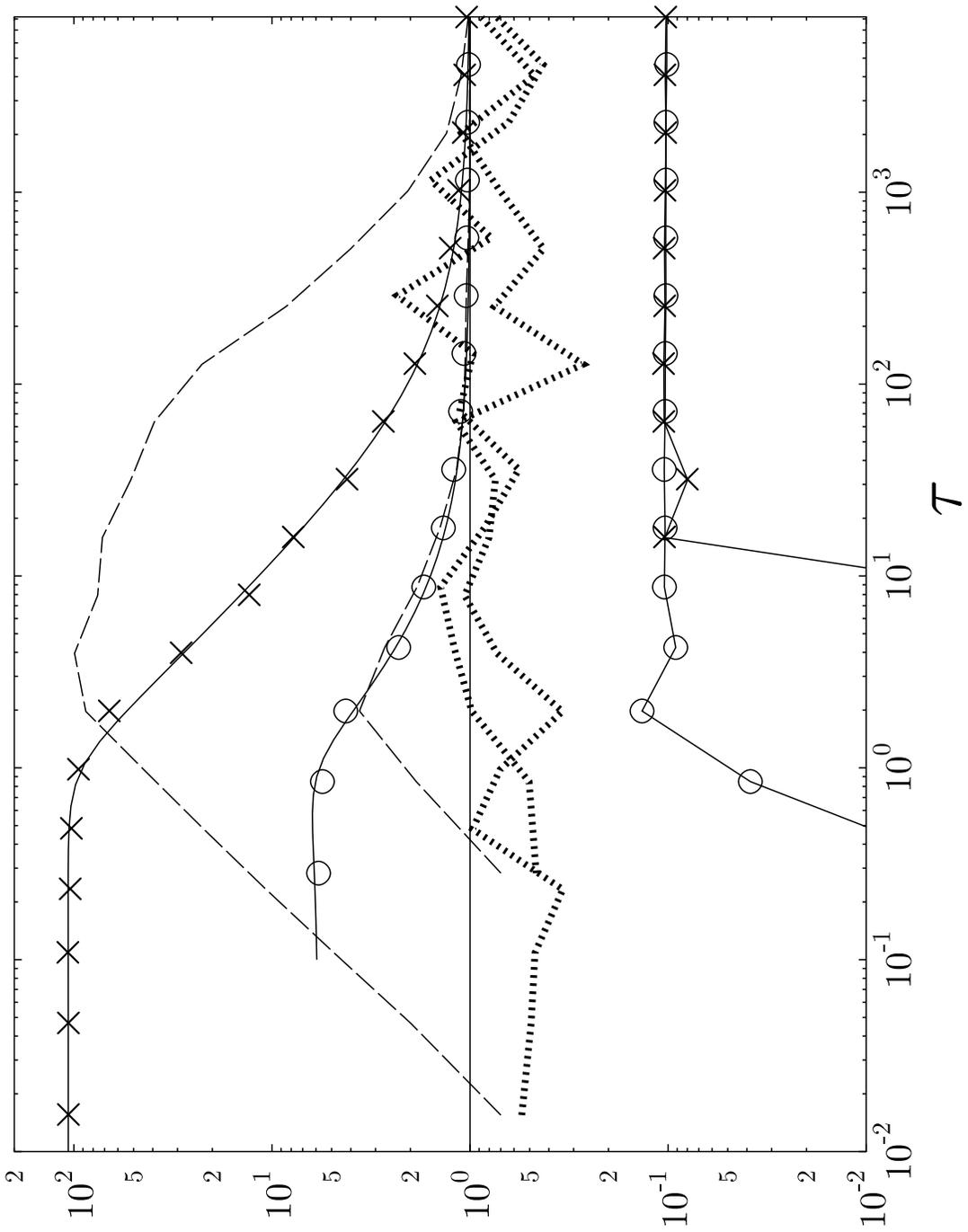
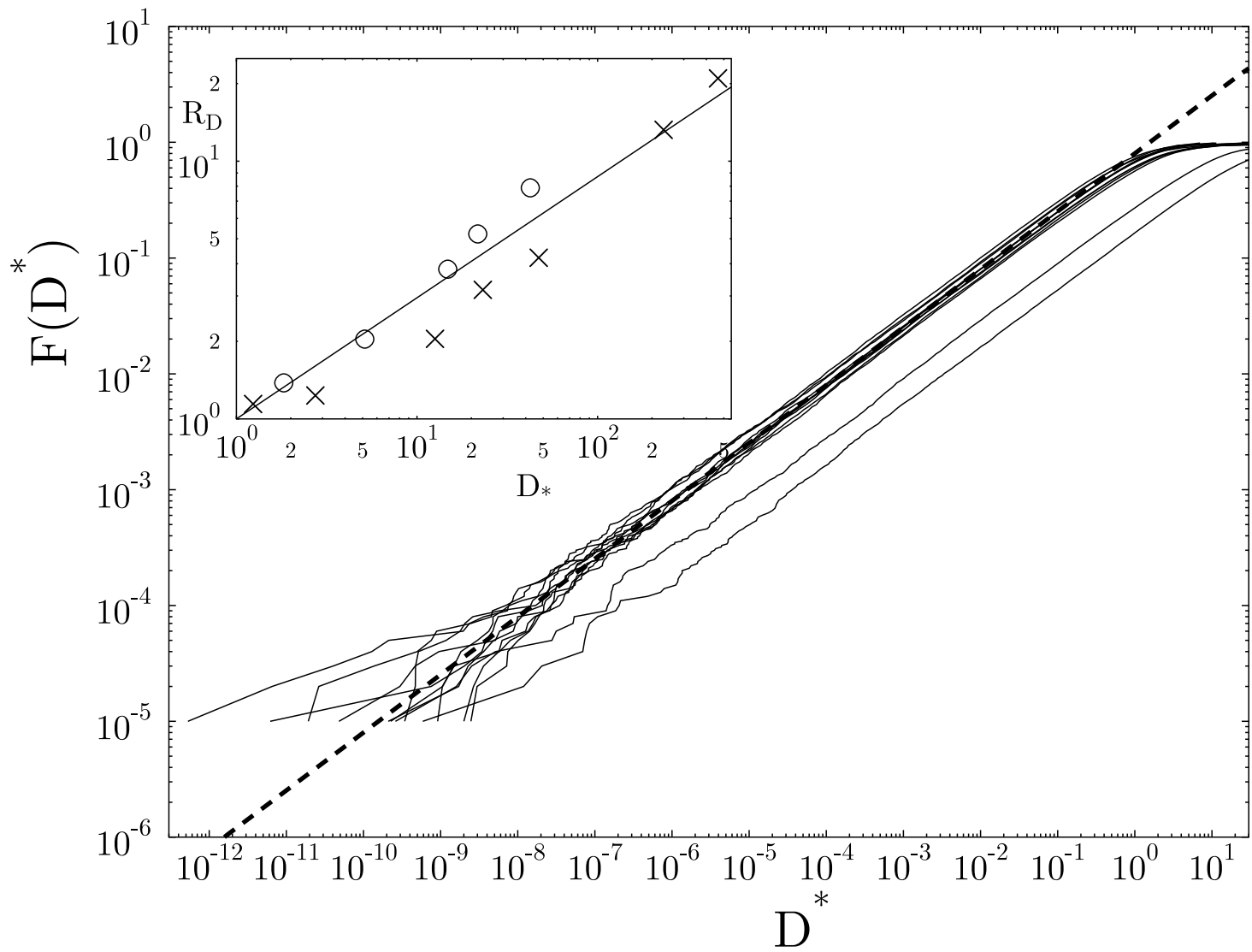
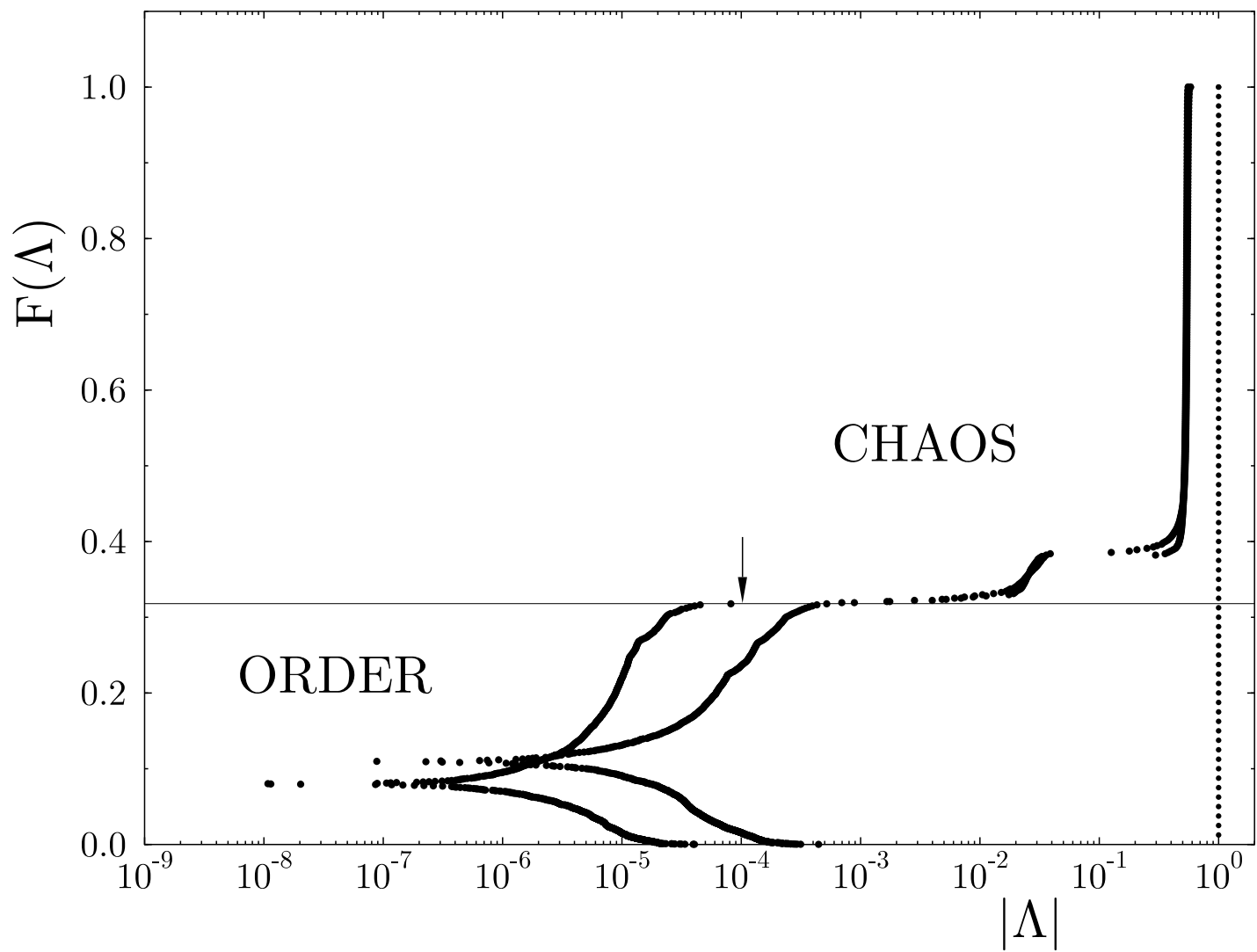
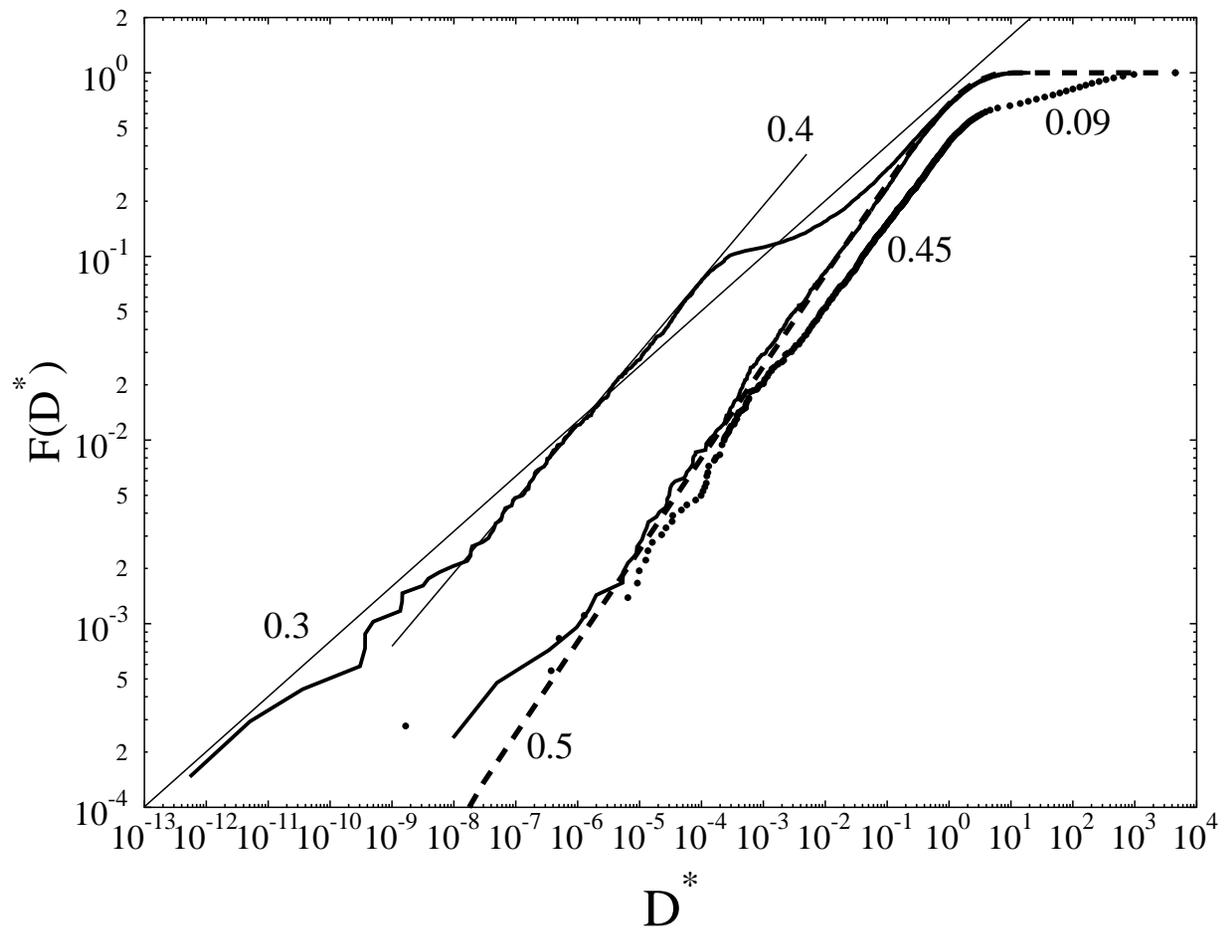
\end{document}